\documentclass{emulateapj}
\usepackage{psfig}

\newcommand{\msun}{M$_{\odot}$}

\shorttitle{Cosmological implications of dwarf spheroidals}
\shortauthors{Fenner et al.}

\begin{document}

\title{Cosmological implications of dwarf spheroidal chemical
evolution}

\author{Yeshe Fenner}
\affil{Harvard-Smithsonian Center for
Astrophysics, 60 Garden Street, Cambridge, MA, 02138, USA}
\email{yfenner@cfa.harvard.edu}

\author{Brad K. Gibson}
\affil{Centre for Astrophysics, University of Central Lancashire,
Preston, PR1 2HE, UK}
\email{bkgibson@uclan.ac.uk}

\author{Roberto Gallino}
\affil{Dipartimento di Fisica Generale, Universita di Torino, 10125
Torino, Italy}
\email{gallino@ph.unito.it }

\and 

\author{Maria Lugaro} 
\affil{Astronomical Institute, University of
Utrecht Postbus 80000, 3508 TA Utrecht, The Netherlands}
\email{M.Lugaro@phys.uu.nl }

\begin{abstract}  

The chemical properties of dwarf spheroidals in the local group are
shown to be inconsistent with star formation being truncated after the
reionization epoch ($z$\,$\sim$\,8). Enhanced levels of [Ba/Y] in
stars in dwarf spheroidals like Sculptor indicate strong $s$-process
production from low-mass stars whose lifetimes are comparable with the
duration of the pre-reionization epoch. The chemical evolution of
Sculptor is followed using a model with SNe~II and SNe~Ia feedback and
mass- and metallicity-dependent nucleosynthetic yields for elements
from H to Pb. We are unable to reproduce the Ba/Y ratio unless stars
formed over an interval long enough for the low-mass stars to pollute
the interstellar medium with $s$-elements. This robust result
challenges the suggestion that most of the local group dwarf
spheroidals are fossils of reionization and supports the case for
large initial dark matter halos.

\end{abstract} 

\keywords{galaxies: abundance ratios --- galaxies: dwarf --- galaxies:
Local Group --- galaxies: evolution --- stars: abundances --- stars:
AGB --- nucleosynthesis}

\section{Introduction}\label{s:intro}

Dwarf galaxies in the Local Group provide a nearby laboratory for
testing dark matter (DM) structure formation scenarios and indirectly
probing the epoch of reionization. Many properties of Milky Way's
dwarf spheroidal (dSph) satellites have been well studied, yet their
role in the cosmological context is still uncertain. Cosmological
simulations have long been plagued by the ``missing satellites''
problem - the overprediction by up to two orders of magnitude of the
abundance of DM subhalos with mass $\lesssim 10^8$\,{\msun} in systems
like our Local Group (Klypin et al. 1999). The two most popular
explanations for the observed deficit of low-mass satellites are that:
1) only massive ($\gtrsim 10^9$\,{\msun}) satellites were able to form
stars, with the spheroidal morphology and small present-day stellar
velocities of dSphs arising through tidal interactions (e.g. Kravtsov
et al. 2004; Kazantzidis et al. 2004); or 2) low mass satellites
($\lesssim 10^8$\,{\msun}) could form stars, but only prior to the
reionization epoch, after which UV heating photoevaporates the gas and
curtails star formation (Barkana \& Loeb 1999; Bullock et al. 2000;
Ricotti \& Gnedin 2005; Gnedin \& Kravtsov 2006). These two solutions
have very different, and testable, consequences for the dSph
population. The first scenario implies that Galactic dSphs descended
from far more massive DM halos than one infers assuming that mass
follows light (i.e. $10^{7-8}$\,{\msun}, Mateo 1998), while the second
scenario implies that many dSphs are very ancient ($z$\,$\gtrsim$\,8)
and associated with less massive DM halos.

Recent work from Ricotti \& Gnedin (2005) favors the second scenario,
in which most dSphs are ``fossils'' of reionization. They suggest that
internal ionizing sources and, to a lesser extent, UV background
radiation prevented the gas from cooling and collapsing to form stars
after the reionization era. In this case, almost all star formation
(SF) activity transpired in an interval of
$\lesssim$\,600\,Myr\footnote{here we have assumed $\lambda$CDM
cosmology with present-day $\omega _{m} = 0.34$, $\omega _{\lambda} =
0.66$, Hubble parameter = 70 km s$^{-1}$ Mpc$^{-1}$, and a 13 Gyr age
for the universe}. This relatively brief window of opportunity for SF
should leave observable signatures on the stellar chemical
properties. A commonly used ``clock'' of SF is $\alpha$/Fe versus
metallicity (Matteucci \& Recchi 2001). This is because
$\alpha$-elements like O, Mg and Si are generated primarily in
short-lived massive stars culminating as Type II supernovae (SNe~II),
whereas a substantial fraction of Galactic Fe is attributed to Type Ia
SNe (SNe~Ia), which occur on longer timescales (Timmes, Woosley \&
Weaver 1995). The first generation of SNe~II pollute their host galaxy
with supersolar $\alpha$/Fe gas, while a delayed contribution from
SNe~Ia subsequently leads to a decrease in $\alpha$/Fe with increasing
Fe/H. Thus, a system with low $\alpha$/Fe at low metallicity is
interpreted as having undergone slow and protracted SF, whereas stars
born in a short intense burst should exhibit higher
$\alpha$/Fe. Unfortunately $\alpha$/Fe does not uniquely trace
SF. Dwarf galaxies are especially vulnerable to feedback effects that
may foster the preferential loss of elements from massive SNe~II, thus
complicating the interpretation of the $\alpha$/Fe diagnostic.

In this paper we employ the abundance pattern of neutron-capture
elements, particularly those associated with the slow neutron-capture
process (known as the $s$-process), as a better chemical test of SF
duration in dSphs. Venn et al. (2004) suggested that the overabundance
of [Ba/Y] in most dSphs was consistent with the production of
$s$-elements from lower-mass stars. We confirm that the curious
behavior of the relative abundance of neutron-capture elements Y, Ba
and Eu in dSphs reflects a strong contribution from low-mass stars,
which are the chief source of the $s$-elements (Busso et
al. 1999). Owing to the correspondingly long lifetimes of these
low-mass stars, one may use chemical evolution simulations to place
limits on the minimum duration of star formation allowed by the
observations.  In Section~\ref{s:model} we give the details of our
chemical evolution model of the representative dwarf galaxy
Sculptor. Section~\ref{s:results} presents results from models with
different star formation histories, in order to assess whether dSphs
are survivors or fossils of the epoch of reionization and
Section~\ref{s:discussion} contains our main conclusions.

\section{The model}\label{s:model}

We have modeled the chemical evolution of the dwarf spheroidal
Sculptor on the assumption that its evolution and SF history is
typical of the types of dSph galaxies that Ricotti \& Gnedin (2005)
identified as likely fossils of reionization. Moreover, its proximity
makes Sculptor one of the best studied dSphs with a large set of
chemical abundance constraints (e.g. Shetrone et al. 2003; Geisler et
al. 2005). This section summarizes the key features of the model,
which is described in greater detail in Fenner et al. (2006, in prep),
where it is applied to ten local group dwarf spheroidals and
irregulars.

\subsection{Equations}\label{s:model:eqns}

The chemical enrichment history of Sculptor was simulated using a
modified version of the chemical evolution code described in Fenner \&
Gibson (2003) and Fenner et al. (2003, 2004, 2005). Major changes made
to the code include: allowing for SNe-driven outflows; employing the
star formation history from Dolphin et al. (2005); and adding s- and
$r$-process yields. The evolution of the gas phase abundance pattern
reflects the cumulative history of the dynamic processes of star
formation, gas inflows and outflows, stellar evolution, and
nucleosynthesis. The set of equations governing these processes were
numerically solved by defining $\sigma_i(t)$ as the mass surface
density of species $i$ at time $t$, and assuming that the rate of
change of $\sigma_i(t)$ is given by:

\begin{eqnarray}\label{eqngce}
\displaystyle\frac{d}{dt} \sigma_{i} (t) & = & E_{i,LIMS}(t) + E_{i,SNII}(t) + E_{i,SNIa}(t)  \nonumber \\
 & & -W_{i,ISM}(t) - W_{i,SNII}(t) - W_{i,SNIa}(t) \nonumber \\
  & & + \displaystyle\frac{d}{dt} \sigma_i(t)_{infall} -X_i(t)\,\psi(t),
\end{eqnarray}

\noindent
where the first three terms, $E_{i,LIMS}(t), E_{i,SNII}(t)$ and
$E_{i,SNIa}(t)$, denote the mass surface density of species $i$
ejected at time $t$ by low- and intermediate-mass stars (LIMS), SNe~II
and SNe~Ia, respectively. The following term, $W_{i,ISM}(t)$, denotes
the mass surface density of species $i$ that escapes from the neutral
galactic interstellar (ISM) due to being carried along with SNe-driven
winds. The next two terms, $W_{i,SNII}(t)$ and $W_{i,SNIa}(t)$,
represent mass of species $i$ released by SNe~II and SNe~Ia,
respectively, that is expelled via winds. The term $\frac{d}{dt}
\sigma_i(t)_{infall}$ gives the infall rate of $i$, which is
determined by calculating the amount of gas required to ensure the SFR
satisfies the Kennicutt law (see Section~\ref{s:model:sfr}).  The
chemical composition of the infalling gas (and the gas at the first
timestep) was taken to be primordial. The final term,
$X_i(t)\,\psi(t)$, gives the depletion of species $i$ due to
incorporation into newly formed stars, where $X_i(t)$ is the mass
fraction of $i$ in the ISM at time $t$ and $\psi(t)$ is the SFR. The
ejection rates in Eqn~\ref{eqngce}, specifying the production of each
element from the different types of star, are given by:

\vspace{0.3cm}

\noindent
$
E_{i,LIMS}(t) = \displaystyle\int^{m_{bl}}_{m_{low}}\psi(t-\tau_{m})\,Y_i(m,Z(t-\tau_{m}))
 \,\frac{\displaystyle\phi(m)}{m} \,\; dm 
$

\begin{equation}
 + \,  (1-\textrm{A})  \displaystyle\int^{m_{bu}}_{m_{bl}}   \psi(t-\tau_{m})\,Y_i(m,Z(t-\tau_{m}))
 \,\frac{\displaystyle\phi(m)}{m} \,\; dm 
\end{equation}

\vspace{0.3cm}

\noindent
$
E_{i,SNIa}(t) =   \textrm{A} \displaystyle\int^{m_{bu}}_{m_{bl}} \frac{\displaystyle\phi(m_{b})}{m}\, \times   
$

\begin{equation}
\left[ \displaystyle\int^{0.5}_{\mu_{m}}f(\mu) 
 \psi(t-\tau_{m_2})\,Y_{b,i}(m,Z(t-\tau_{m_2})) d\mu \right]
  \,\; dm_b
\end{equation}

\begin{equation}
E_{i,SNII}(t) = \displaystyle\int^{m_{up}}_{m_{bu}}\psi(t-\tau_{m})\,Y_i(m,Z(t-\tau_{m}))
 \,\frac{\displaystyle\phi(m)}{m} \,\; dm,  
\end{equation}

\vspace{0.2cm}

\noindent
where $Y_i(m,Z(t-\tau_{m}))$ is the stellar yield of $i$ (in mass
units) from a star of mass $m$, main-sequence lifetime $\tau_{m}$, and
metallicity $Z(t-\tau_{m})$. The initial mass function, $\phi(m)$, has
upper and lower mass limits of $m_{up}$=60\,{\msun} and
$m_{low}$=0.08\,{\msun}, respectively.  The models presented here
employ the Kroupa, Tout \& Gilmore (1993) three-component IMF.  The
fraction of LIMS in binary systems culminating in a SNe~Ia is given by
the parameter A. The SNe~Ia progenitor binary mass is denoted by $m_b$
and is the sum of the primary and secondary masses, $m_1 + m_2$. The
binary mass has upper and lower limits of $m_{bu}$=16\,{\msun} and
$m_{bl}$=3\,{\msun}, respectively. The function $f(\mu)$ determines
the distribution for the mass fraction of the secondary ($\mu$ =
$m_2/m_b$). The reader is referred to Matteucci \& Greggio (1986) for
more detailed discussion and definitions of the terms in the above
equations.

Most previous studies (e.g. Lanfranchi et al. 2003; 2004; 2006;
Robertson et al. 2005) have assumed differential galactic winds in
proportion to the SFR. These have assumed $W_{i}(t) = w_i \psi(t)$,
where $w_i$ is a free parameter that is arbitrarily set for each
element in order to reproduce observations, and is then kept constant
as a function of time. Such a scheme does not allow for behavior such
as increased escape of Fe relative to O after a burst of SF when the
frequency of SNe~Ia may compare with, or exceed, that of
SNe~II. Instead, we self-consistently calculate the amount of SNe~II and
SNe~Ia ejecta lost at each timestep as a function of the appropriate
SNe rates. Cold interstellar gas also escapes from the galaxy by
imposing a mass-loading factor that is consistent with
observations. The form of the equations governing the loss of species
$i$ in galactic winds can be written:

\begin{equation}
W_{i,SNIa}(t) = E_{i,SNIa}(t) \times \textrm{min}(0.9,\frac{\varepsilon_{SNIa}}{m_{tot}} \times R_{SNIa}) 
\end{equation}

\begin{equation}
W_{i,SNII}(t) =  E_{i,SNII}(t) \times \textrm{min}(0.9,\frac{\varepsilon_{SNII}}{m_{tot}} \times R_{SNII})
\end{equation}

\begin{equation}
W_{i,ISM}(t) =  X_i(t) \times \textrm{ML} \times (W_{gas,SNIa}(t) + W_{gas,SNII}(t))
\end{equation}

\vspace{0.2cm}

\noindent
where $m_{tot}$ is the total galaxy mass in units of $10^6$\,{\msun}
(taken from Mateo 1998). The Type Ia and II SNe rates are given by
$R_{SNIa}$ and $R_{SNII}$, while the corresponding wind efficiency
factors are $\varepsilon_{SNIa}$ and $\varepsilon_{SNII}$. The mass of
element $i$ in the ISM that gets entrained in the SNe-driven winds and
driven out of the galaxy is set by the mass-loading factor, ML
($=$\,15, as discussed below). The sum of $W_{gas,SNIa}(t)$ and
$W_{gas,SNII}(t)$ gives the total amount of stellar ejecta lost in
winds at time $t$, and is equivalent to the sum of $W_{i,SNIa}(t)$ and
$W_{i,SNII}(t)$ over all $i$.  The escape fraction of SNe~II and SNe~Ia
ejecta increases linearly with their respective SNe rates, but the
overall efficiency is inversely proportional to the depth of the
galactic potential well, as measured by the total galaxy mass,
$m_{tot}$. We do not allow more than 90\% of the SNe ejecta to escape
at each timestep, based on the prediction by Hensler et al. (2004)
that up to 25\% of SNe material gets mixed into the local ISM on short
timescales rather than being carried away by an expanding
superbubble. We note that the main conclusions of this paper are
robust to the implementation of feedback.

Martin et al. (2002) estimated that the diffuse halo gas of the dwarf
starburst NGC~1569 has $\alpha$/Fe $\sim$ 2-4 times solar. The
metallicity of the hot halo gas was measured to be roughly solar,
implying a mass-loading factor of $\sim$\,9. In order to obtain [O/Fe]
$\sim$ +0.3 to +0.6 in the galactic winds, we assume that SNe~Ia and
SNe~II have the same explosion energy of 10$^{51}$\,erg and we set the
feedback efficiency for SNe~Ia to be five times higher than that of
SNe~II. This might seem counter intuitive, given that SNe~II occur in
associations, which should enhance their ability to heat the ISM and
create bubbles and chimneys facilitating the escape of
metals. However, Recchi et al. (2004, 2006) argue that SNe~II are
\emph{less} efficient at expelling their ejecta, because much of their
energy goes toward heating the cold dense molecular clouds from which
they were born. The characteristic lifetimes of SNe~Ia range from tens
to thousands of Myr, by which time the SNe~Ia progenitors will have
traveled to regions of lower density, where less energy is required to
widely disperse ejecta.

We assume that turbulent motions at the boundary between the
SNe-driven winds and the ambient ISM mix significant quantities of
cold gas into the outflows. Our adopted mass-loading factor of 15 is
similar to the value inferred for NGC~1569 and the value of 10
calculated by Silk (2003). The mass-loading factor (i.e. the mass of
the ISM carried away with the winds relative to the mass of stellar
ejecta in the wind) is the key parameter determining the metallicity
of the galactic outflows.

Observations of superwinds in starburst galaxies by Heckman et
al. (2000) reveal outflow rates (OFR) comparable with the SFR.
Theoretical calculations support this approximate equivalence between
the OFR and SFR (Silk 2003; Shu, Mo \& Mao 2005). Silk (2003)
estimated that OFR $\sim$ SFR, in the case of large porosity, and OFR
$<$ SFR, when porosity is low. In accordance, our overall wind
efficiency was chosen to produce outflow rates roughly half the
SFR. We also assumed that galactic wind efficiency is inversely
proportional to the total galaxy mass, where galaxy mass was taken
from Mateo (1998). 

Given a velocity dispersion of about 2 km/s for stars in OB
associations (Tian et al. 1996) and lifetimes $>$\,60\,Myr for low-
and intermediate-mass stars (Schaller et al. 1992), then most LIMS
will have have drifted outside the $\sim$ 50-100 pc radii of the
stellar association in which they were born by the time they begin to
pollute the ISM with their winds. Owing to their weak stellar winds
and their likely migration from OB bubbles, we mix the newly released
ejecta from LIMS directly into the ISM at each timestep.

\subsection{Star Formation History (SFH)}\label{s:model:sfr}

\begin{figure}
\plotone{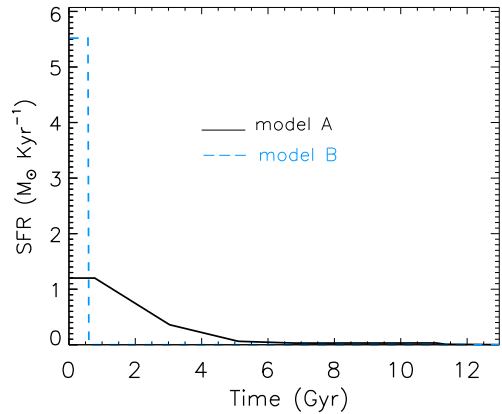}
      \caption{Star formation rate in units of {\msun} per thousand
      years as a function of time for models A and B. }\label{fig:sfr}
\end{figure}

We tested two different star formation histories: 1) model A adopts
the SF history inferred from Sculptor's color-magnitude diagram
(Dolphin et al. 2005) in which SF extends for over several Gyrs; 2)
model B compresses the SF within a period of
600\,Myr. Fig.~\ref{fig:sfr} compares these two SF histories.  The gas
accretion rate is often treated as an input into chemical evolution
models, however we work backwards to infer the infall rates from the
SFHs by inverting the Kennicutt (1998) SF law. Like Lanfranchi et
al. (2003, 2004, 2006), we find that low SF efficiencies are better
able to reproduce the chemical abundances in local group dwarf
galaxies. In particular, we obtain good fits in our standard model
using $\psi(t) = 0.05 \times
\sigma(t)^{1.4}$\,M$_{\odot}$\,pc$^{-2}$\,Gyr$^{-1}$, where $\psi$ is
the star formation rate (SFR) and $\sigma$ is the gas surface
density. The difference between the $\sigma(t)$ implied by the
$\psi(t)$ and the value predicted within the chemical evolution
models, provides the infall rate at each timestep.

\subsection{Stellar Yields}\label{s:model:yields}

\subsubsection*{Low and intermediate mass stars (LIMS)} 

For the production of isotopes up to the Fe-peak from stars less
massive than 8 {\msun}, we incorporated yields from the stellar
evolution and nucleosynthesis code described in Karakas \& Lattanzio
(2003), supplemented with unpublished yields for the metallicity
$Z$=0.0001.  The origin and interpretation of $s$-process elements
like La, Ba and Y in the local group is of considerable import to this
study. The Karakas et al. models do not calculate the $s$-process, so
for elements above the Fe-peak, we employ the nucleosynthetic yields
of the Torino group (Gallino et al. in prep). These mass and
metallicity dependent yields have previously been used in the Galactic
chemical evolution models of Travaglio et al. (1999, 2001, 2004). In a
similar fashion to Travaglio et al., we have averaged the $s$-process
yields over a range of $^{13}$C pocket choices.

The slow neutron-capture process in asymptotic giant branch (AGB)
stars is thought to be driven primarily by the
$^{13}$C$(\alpha,n)^{16}$O reaction, which provides the flux of
neutrons that can be captured by elements from the Fe-peak upwards
(Busso et al. 1999). The source of the $^{13}$C probably comes from
CN-cycling in the H envelope leading to proton capture by $^{12}$C and
the development of a pocket of $^{13}$C at the top of the intershell
region between the H-shell and He-shell. This pocket arises through
the diffusion of a small amount of protons from the H envelope at the
epoch of the third dredge-up episodes. Proton mixing is induced by the
close contact between the convective H-rich envelope and the radiative
He-rich and C-rich He intershell, this forming a tiny "proton pocket".
These protons are subsequently captured by the abundant $^{12}$C in
the He intershell, creating a $^{13}$C pocket of primary nature.
Afterwards, when the pocket is heated up to about 0.9 $\times
10^8$\,K, $^{13}$C is consumed by $\alpha$ captures that release
neutrons and give rise to a very intense neutron exposure. The
formation of the $^{13}$C pocket is not simulated within the
nucleosynthesis codes, but rather is treated parametrically and
constrained by observations (Busso et al. 1999).

\subsubsection*{Type~Ia supernovae (SNe~Ia)} 

We adopted a recalculation of the Thielemann, Nomoto, \& Yokoi (1986)
W7 model by Iwamoto et~al. (1999) to estimate the yields from SNe~Ia.
It was assumed that 4\% of binary systems involving intermediate and
low mass stars result in SNe~Ia, since this fraction provides a good
fit to the solar neighborhood (e.g.  Alib{\' e}s, Labay, \& Canal
2001; Fenner \& Gibson 2003).

\subsubsection*{Massive stars} 

For stars more massive than 8--10~{\msun} that end their lives in
violent supernova explosions, we implemented the yields from Woosley
\& Weaver (1995) for elements up to Zn (using same technique as Fenner
et al. 2003, 2004 and halving the Fe yield, as suggested by Timmes,
Woosley, \& Weaver 1995). Beyond Zn, the yields of $r$-process
elements like Eu are difficult to calculate within nucleosynthesis
models and it is common practice to assume primary production from
SNe~II and empirically deduce them (e.g. Argast et al. 2004) or
estimate them by calculating the difference between the solar
abundance and the pure $s$-process contribution predicted by Galactic
chemical evolution models (e.g. Travaglio et al. 1999, 2001;
2004). Although the stellar mass range responsible for r-production is
still under debate (e.g. Sumiyoshi et al. 2001; Wanajo et al. 2003;
Tsujimoto et al. 2000), our models link the $r$-process site to all
SNe~II with mass $<$\,40\,M$_{\odot}$. We tested the case in which only
SNe~II with mass $<$\,15\,M$_{\odot}$ produced r-elements, finding that
the conclusions of this paper are not very sensitive to the adopted
mass range.  Metallicity-dependent main-sequence lifetimes calculated
by Schaller et~al. (1992) have been employed.


\section{Results}\label{s:results}

\begin{figure}
\hspace{-0.5cm}\epsscale{1.2}\plotone{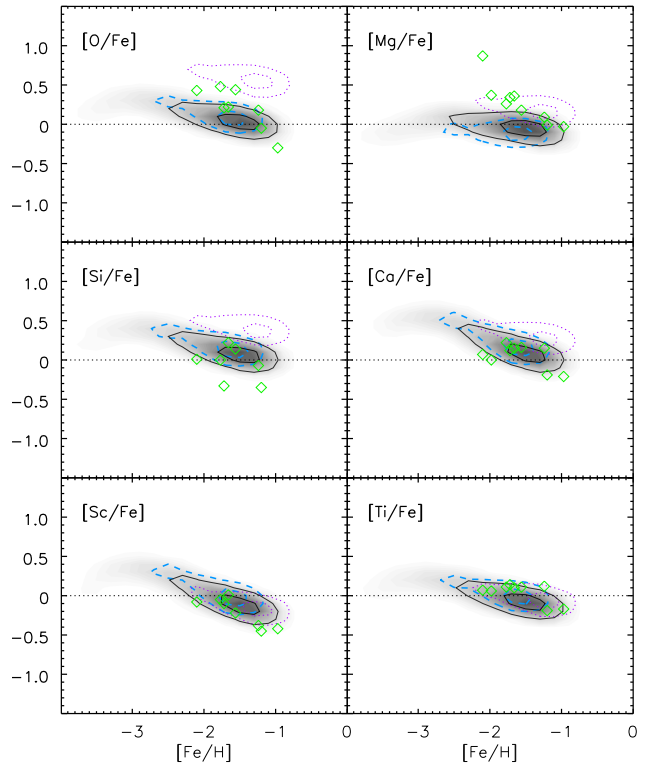}
      \caption{Predicted abundances for the Sculptor dSph chemical
      evolution model, incorporating SN feedback and galactic winds,
      plotted against stellar observations from Shetrone et al. 2003
      and Geisler et al. 2005 (\emph{diamonds}).  The shaded region
      indicates the relative frequency of stars from model A as a
      function of [X/Y] and [Fe/H], where [X/Y] is indicated in the
      top lefthand corner of each panel. Pairs of contour lines are
      plotted at 0.25 and 0.75 of the maximum frequency. The predicted
      stellar abundances have been convolved with a Gaussian of
      dispersion 0.1\,dex to mimic observational
      uncertainties. Results from model B, in which SF is shut-off by
      reionization after 0.6\,Gyr, are shown by dashed contour
      lines. Dotted contour lines correspond to model C, which is the
      same as model A but without SNe feedback.}\label{fig:alpha}
\end{figure}

\begin{figure}
\hspace{-0.5cm}\epsscale{1.2}\plotone{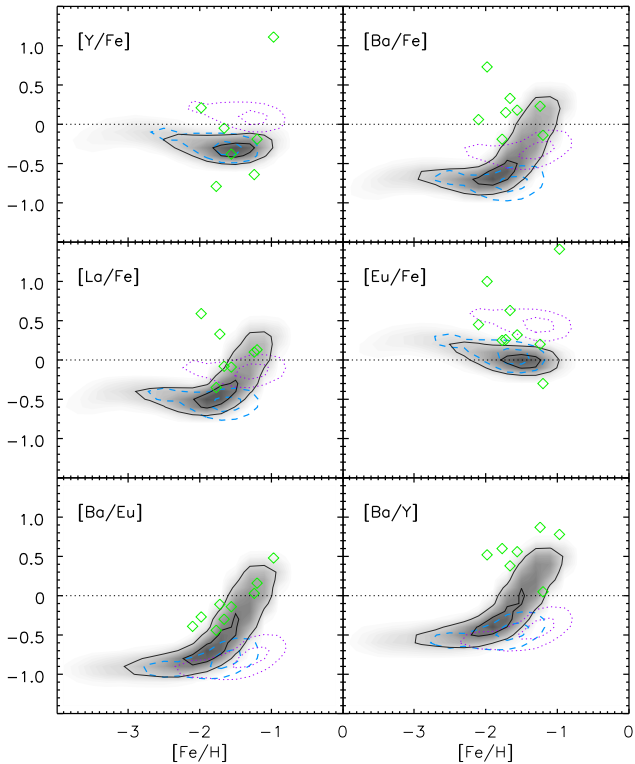}
      \caption{Same as Fig.~\ref{fig:alpha} but for neutron-capture
      elements. The shaded region indicates the relative frequency of
      stars from model A as a function of [X/Y] and [Fe/H], where
      [X/Y] is indicated in the top lefthand corner of each
      panel. Pairs of contour lines are plotted at 0.25 and 0.75 of
      the maximum frequency. The predicted stellar abundances have
      been convolved with a Gaussian of dispersion 0.1\,dex to mimic
      observational uncertainties. Results from model B, in which SF
      is shut-off by reionization after 0.6\,Gyr, are shown by dashed
      contour lines. Dotted contour lines correspond to model
      C.}\label{fig:ncapt}
\end{figure}

Fig.~\ref{fig:alpha} shows the predicted abundances of O, Mg, Si, Ca,
Sc and Ti relative to Fe from the Sculptor model using: i) the
standard SFH inferred from observations (model A: solid contour lines
and shaded region); and ii) the suppressed SF model (model B: dashed
contour lines). To illustrate the role of feedback and preferential
loss of SNe ejecta, we also show results from an identical simulation
to model A but \emph{without} galactic winds (model C: dotted contour
lines). Red giant stars observed in Sculptor by Shetrone et al. (2003)
and Geisler et al. (2005) (diamonds) have a mean
[Fe/H]\,$\sim$\,$-$1.57, which is well matched by our predicted mean
of [Fe/H]\,$\sim$\,$-$1.65. We also produce a metallicity spread in
agreement with the data. Sculptor stars exhibit $\alpha$/Fe
ratios\footnote{where $\alpha$-elements include O, Mg, Si, Ca and
Ti. Although we also plot Sc in this figure, it is classed as an
iron-peak element.}  slightly above solar at the lowest metallicities,
before declining to subsolar values at higher [Fe/H]
($\gtrsim$\,$-$1.5). Note that the subsolar [$\alpha$/Fe] values in
Sculptor stars contrasts dramatically with the supersolar ratios
characterizing MW stars at the same metallicity. This is thought to be
common in dwarf galaxies whose low gas surface densities lead to
inefficient SF and consequently long metal enrichment
timescales. However, we find that subsolar $\alpha$/Fe can arise
through the preferential loss of SN products via galactic winds
(compare models A and C). Both the standard and suppressed SF models
satisfactorily reproduce the overall $\alpha$-element trends. The
standard SF model has slightly lower (by\,$\sim$\,0.1\,dex) O, Si, Ca,
Sc and Ti with respect to Fe than in the suppressed SF model, but the
influence of galactic winds makes the difference smaller than for a
closed box model. The prevention of SNe feedback (dotted contour
lines) leads to an excess of O, Si and Ca, in conflict with the data.
Fig.~\ref{fig:alpha} demonstrates that [$\alpha$/Fe] is not a clean
probe of SF history in galaxies that have experienced metal-enriched
outflows. Indeed, [$\alpha$/Fe] is more sensitive in our Sculptor
models to SNe feedback than to the duration of SF.

Conversely, Fig.~\ref{fig:ncapt} suggests that the neutron-capture
elements provide a more useful discriminant of the different star
formation models. The neutron-capture elements show curious patterns
in Sculptor stars (e.g. Venn et al. 2004).  The light $s$-process
element Y tends to be subsolar relative to Fe, and lower than in Milky
Way stars of comparable metallicity. Conversely, the heavy $s$-process
element Ba is slightly higher than in Galactic field stars. In our
models, the Sculptor stars with enhanced Ba/Y ratios were born from
gas that had been heavily polluted by low-mass stars. The high [Ba/Y]
ratio can only be matched by our model incorporating both SNe-driven
winds and SF extending over several Gyr. This is because the origin of
gas with [Ba/Y]\,$\gtrsim$\,+0.5\,dex is low-mass metal-poor stars on
the asymptotic giant branch, whose weak ($\sim$\,15--30\,km\,s$^{-1}$)
winds are easily retained by the potential well.

For low-mass AGB stars to dominate $s$-element abundances in the ISM
there needs to be substantial loss of SNe~II ejecta \emph{and} SF
timescales exceeding the lifetimes of low-mass stars. The strong
metallicity-dependence of $s$-process yields from AGBs gives rise to
the sharp upturn in Ba and La abundance at
[Fe/H]\,$\sim$\,$-$1.6. This leads to some stars having
Ba-enhancements as high as observed, although the fraction of
simulated stars with supersolar Ba relative to Fe, Y and Eu is
underestimated, even in model A. This may inform us about the size and
nature of the $^{13}$C pocket in metal-poor AGB stars.  AGB stars
produce heavier $s$-elements like La and Ba more efficiently than
lighter species like Y at low metallicities. This is because the
neutron flux per seed nuclei varies roughly inversely with
metallicity. Since the slow neutron-capture process proceeds further
along the periodic table with decreasing initial stellar metallicity,
Pb production from AGBs peaks at lower [Fe/H], followed by a Ba and La
peak at higher [Fe/H]. The production of Y from AGBs reaches peak
efficiency at higher [Fe/H] than is typical for dSphs. A further
prediction from our model is that dwarf spheroidal stars should have
supersolar [Pb/Ba] and [Pb/Fe] ratios. Although Pb requires very high
signal-to-noise and has not yet been measured in stars outside the
Milky Way, sufficient exposure time may yield future detections in red
giants in nearby dSphs. Such measurements would be a very useful test
of this model and may provide insight into the nature of the
$s$-process in low-mass stars.

The relative importance of the $s$- and $r$-process is usually
measured by the [Ba/Eu] ratio, where the pure $r$-process ratio is
[Ba/Eu]\,$\sim$\,$-$0.8\,dex and the pure $s$-process ratio is
$\sim$\,+1\,dex (Arlandini et al. 1999).  Fig.~\ref{fig:ncapt} shows
about an order of magnitude rise in Ba/Eu going from [Fe/H]\,=\,$-$2
to $-$1. Sculptor stars tend towards [Ba/Eu]\,$=$\,$+$0.5 at
[Fe/H]\,$=$\,$-$1, which is 0.8\,dex above the corresponding MW value.
Only Model A reproduces this trend, owing to the retention of
lower-mass stellar ejecta and the preferential loss of the ejecta from
SNe~II, which are major suppliers of Eu. The good agreement between
Model A predictions and the observed [Ba/Fe], [La/Fe] and [Ba/Eu],
lends further support to the case for extended SF in dSphs. We place
more weight on the Ba/Y diagnostic, however, since there is no obvious
alternative to low-mass stars that can match the high values measured.

Interestingly, the massive Galactic globular cluster $\omega$ Centauri
exhibits similar chemical properties, with the lowest metallicity
stars ([Fe/H\,$\sim$\,$-$1.8) having a predominantly $r$-process
abundance pattern and those with the highest metallicity
([Fe/H]\,$\sim$\,$-$0.8) having dramatically enhancements in
$s$-process elements (Smith et al. 2000). Smith et al. attributed
these observations to the greater retention of low-mass stellar ejecta
- because of their weak winds - compared with energetic SNe~II
winds. They infer that star formation must have continued for at least
several Gyr, making $\omega$ Centauri highly anomalous compared with
other globular clusters, whose individual stellar populations formed
contemporaneously. Indeed, it has been suggested (e.g. Majewski et
al. 2000) that $\omega$ Centauri is not a globular cluster, but the
remains of an accreted dwarf spheroidal.

\begin{figure}
\hspace{-0.5cm}\epsscale{1.2}\plotone{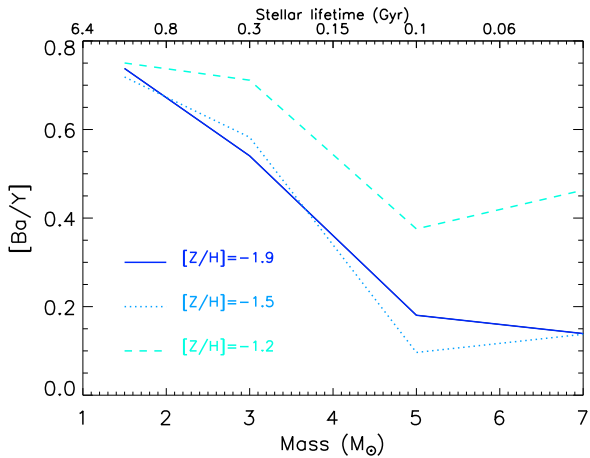}
      \caption{[Ba/Y] in the ejecta from AGB stars versus initial
      stellar mass for the three metallicities indicated (for a choice
      of $^{13}$C pocket mass of $\sim 4 \times 10^{-6}$\,{\msun})
      Corresponding stellar lifetimes from Schaller et al 1992 (for
      Z=0.001) are indicated along the top axis.  }\label{fig:baony}
\end{figure}

To further illustrate the time delay associated with $s$-process
enrichment, Fig.~\ref{fig:baony} displays the ratio of [Ba/Y] in the
ejecta of AGB stars as a function mass for the three metallicities
[Z/H]=$-$1.9, $-$1.6 and $-$1.2 for the standard choice of $^{13}$C
pocket mass (see Travaglio et al. 1999, 2001, 2004 for further details
regarding the $^{13}$C pocket). The corresponding stellar lifetimes
(from Schaller et al. 1992, for a metallicity of Z=0.001) are
presented along the top axis. Sculptor stars have a mean
$<$[Fe/H]$>$\,$\sim$$-$1.6\,dex and
$<$[Ba/Y]$>$\,$\sim$\,0.5\,dex. From Fig.~\ref{fig:baony} it can be
seen that at metallicities typical of dSphs, only AGB stars with mass
$\lesssim$\,3\,{\msun} and lifetimes $\gtrsim$\,300\,Myr have yields
with [Ba/Y]\,$\gtrsim$\,0.5\,dex.  Stars whose yields have [Ba/Y]
equal to the highest values measured in Sculptor stars have lifetimes
$>$\,1\,Gyr. If these stars polluted the ISM with Ba-rich gas from
which subsequent generations of stars formed, then stars must have
been forming for several Gyr or more. We note that with a factor of
1.5 reduction in the $^{13}$C pocket, slightly more massive AGB stars
can yield [Ba/Y]\,$\gtrsim$\,0.5\,dex on shorter timescales (mass
$\lesssim$\,4\,{\msun} and lifetimes $\gtrsim$\,150\,Myr).  Indeed,
smaller $^{13}$C pockets may help resolve the underproduction of
[Ba/Y] and [Ba/Fe] seen in Fig.~\ref{fig:ncapt}. After weighting by
the IMF, however,  lower-mass stars dominate the production of
$s$-elements and still require extended SF in order to pollute the ISM
to the observed levels.

We have argued that the suppression of SF by reionization after
$\sim$\,600\,Myr would not grant low-mass stars enough time to pollute
the ISM with $s$-elements to the extent demanded by the
observations. However, there is still uncertainty regarding the
precise redshift at which the universe was reionized.  To test the
sensitivity of our results to the assumed redshift of reionization, we
ran an identical model with $z_{reion}$\,$\sim$\,6, which is roughly
the lowest value permitted by observations (e.g. Benson et al. 2005,
Fan et al. 2005). This model generously allowed SF to continue for
0.9\,Gyr, yet the mean [Ba/Y] and [Ba/Eu] increased by less than
0.15\,dex and still violated the empirical constraints.

Grebel \& Gallagher (2004) found little evidence for reionization
suppressing local group galaxy evolution, based on: i) metallicity
spreads; ii) low [$\alpha$/Fe] ratios; and iii) stellar ages. However,
a range in metallicites along with low [$\alpha$/Fe] can both arise
from a short star formation episode (see dashed contour lines in
Fig.~\ref{fig:alpha}). The clearest sign of ongoing SF would be an age
spread, however, age-dating old stellar populations is complicated by
the age-metallicity degeneracy and is subject to uncertainties
$\gtrsim$\,1\,Gyr (e.g. Hernandez et al. 2000). Using $s$-element
enrichment from long-lived low-mass stars as an alternative
chronometer, we arrive at the same conclusion as Grebel \& Gallagher
(2004).

If early reionization was not responsible for shutting off SF in dSphs
like Sculptor, what ultimately caused star formation to cease after a
few Gyrs? In our models, SNe feedback can remove the bulk of the
\emph{metals} from Sculptor (and the other dSphs, as shown in Fenner
et al. 2006, in prep) but only a small fraction of the ISM gets
carried away by the SNe-driven winds. Consequently, our Sculptor model
has a final gas fraction of $\sim$\,70\%, whereas most dSphs are
observed to have gas fractions less than a percent.  Lower gas
fractions could be obtained by invoking unreasonably strong galactic
winds, however a more likely mechanism for gas removal is via tidal-
and ram-pressure stripping when the orbits of the dSphs bring them
into close proximity with the Milky Way halo (e.g. Marcolini et
al. 2003, 2004, 2006; Mayer et al. 2005).



\section{Discussion and Conclusions}\label{s:discussion}

We find that the abundance of neutron-capture elements in dwarf
spheroidal stars is indicative of a strong contribution from
long-lived low-mass stars. While this study presented models of
Sculptor, our conclusions apply equally to other old gas-poor dSphs
like Draco, Ursa Minor and Sextans, since they also exhibit enhanced
Ba/Y and have similar star formation histories to Sculptor (Dolphin et
al. 2005).  The observed supersolar Ba/Y ratios imply that stars
formed over an interval of at least several Gyr to allow time for
metal-poor AGB stars to enrich the ISM with $s$-elements. These
results cast doubt on recent suggestions that local group dSphs Draco,
Ursa Minor and Sextans are ``fossils'' of reionization (Ricotti \&
Gnedin 2005; Gnedin \& Kravtsov 2006). Because the Ba/Y ratio is very
high in all dSphs for which it has been measured, we propose that all
dSphs were actively forming stars beyond the epoch of
reionization. This is indirect evidence for their initial dark matter
halos being larger than the $\sim$\,10$^{7}$\,{\msun} values obtained
by assuming that mass follows light (Mateo 1998). 

The case for large mass-to-light ratios and extended DM halos in dSphs
continues to strengthen. Mashchenko et al. (2005) estimated virial
masses of $\sim$\,10$^9$\,{\msun} for Draco, Sculptor and Carina,
based on N-body simulations designed to fit the observed stellar
density and velocity profiles. Dehnen \& King (2006) indirectly infer
a DM halo mass $\gtrsim$\,10$^9$\,{\msun} for Sculptor, from its large
abundance of X-ray binaries, whose high expected velocities should
have facilitated their escape if the potential well was
smaller. Moreover, there appears to be an absence of tidal heating at
large radii in Draco, Ursa Minor, Sextans and Sculptor, which implies
large and extended DM halos (Read et al. 2006; Coleman et al. 2005).
The chemical properties of the representative dwarf spheroidal
Sculptor are well matched by our simulations incorporating SNe-driven
feedback and empirical star formation histories. We contend that the
neutron-capture abundance pattern reflects enrichment from metal-poor
low-mass stars, which is inconsistent with a cessation in star
formation after the reionization epoch and instead supports dSphs
having large initial DM halos.

\vspace{-0.7cm}


\acknowledgements 

Y. F. thanks Brant Robertson, Adam Lidz and Timothy Beers for valuable
conversations and comments.  Y. F. acknowledges the support of an ITC
fellowship from the Harvard College Observatory. M. L. has been
supported by the NWO VENI grant. R.G. acknowledges the support by the
Italian MIUR-FIRB project ``Origin of the heavy elements beyond Fe''.
We thank the referee for useful comments.


\begin{thebibliography}{}

\bibitem[Alib{\' e}s, Labay, \& Canal(2001)]{2001A&A...370.1103A} Alib{\'
e}s, A., Labay, J., \& Canal, R.\ 2001, A\&A, 370, 1103

\bibitem[Argast et al.(2004)]{2004A&A...416..997A} Argast, D., Samland, M., 
Thielemann, F.-K., \& Qian, Y.-Z.\ 2004, A\&A, 416, 997 
 
\bibitem[Arlandini et al.(1999)]{1999ApJ...525..886A} Arlandini, C., 
K{\"a}ppeler, F., Wisshak, K., Gallino, R., Lugaro, M., Busso, M., \& 
Straniero, O.\ 1999, ApJ, 525, 886 
 
\bibitem[Barkana \& Loeb(1999)]{1999ApJ...523...54B} Barkana, R., \& Loeb, 
A.\ 1999, ApJ, 523, 54 

\bibitem[Benson et al.(2005)]{2005astro.ph.12364B} Benson, A.~J., Sugiyama, 
N., Nusser, A., \& Lacey, C.~G.\ 2005, ArXiv Astrophysics e-prints, 
arXiv:astro-ph/0512364 
 
\bibitem[Bullock et al.(2000)]{2000ApJ...539..517B} Bullock, J.~S., 
Kravtsov, A.~V., \& Weinberg, D.~H.\ 2000, ApJ, 539, 517 
 
\bibitem[Busso et al.(1999)]{1999ARA&A..37..239B} Busso, M., Gallino, R., 
\& Wasserburg, G.~J.\ 1999, ARA\&A, 37, 239 
 
\bibitem[Coleman et al.(2005)]{2005AJ....130.1065C} Coleman, M.~G., Da 
Costa, G.~S., \& Bland-Hawthorn, J.\ 2005, AJ, 130, 1065 
 
\bibitem[Dehnen \& King(2006)]{2006MNRAS.tmpL...4D} Dehnen, W., \& King, 
A.\ 2006, MNRAS, L4 
 
\bibitem[Dolphin et al.(2005)]{2005astro.ph..6430D} Dolphin, A.~E., Weisz, 
D.~R., Skillman, E.~D., \& Holtzman, J.~A.\ 2005, ArXiv Astrophysics 
e-prints, arXiv:astro-ph/0506430 
 
\bibitem[Fan et al.(2005)]{2005astro.ph.12082F} Fan, X., et al.\ 2005, 
ArXiv Astrophysics e-prints, arXiv:astro-ph/0512082 
 
\bibitem[Fenner \& Gibson(2003)]{2003PASA...20..189F} Fenner, Y.~\& Gibson, 
B.~K.\ 2003, Publ.~Astron.~Soc.~Aust., 20, 189 

\bibitem[Fenner et al.(2003)]{2003PASA...20..340F} Fenner, Y., Gibson, 
B.~K., Lee, H.-c., Karakas, A.~I., Lattanzio, J.~C., Chieffi, A., Limongi, 
M., \& Yong, D.\ 2003, Publ.~Astron.~Soc.~Aust., 20, 340 

\bibitem[Fenner, Prochaska, \& Gibson(2004)]{2004ApJ...606..116F} Fenner, 
Y., Prochaska, J.~X., \& Gibson, B.~K.\ 2004, ApJ, 606, 116 

\bibitem[Fenner et al.(2004)]{2004MNRAS.tmp..280F} Fenner, Y., Campbell, 
S., Karakas, A.~I., Lattanzio, J.~C., \& Gibson, B.~K.\ 2004, MNRAS, 280 

\bibitem[Geisler et al.(2005)]{2005AJ....129.1428G} Geisler, D., Smith, 
V.~V., Wallerstein, G., Gonzalez, G., \& Charbonnel, C.\ 2005, AJ, 129, 
1428 
 
\bibitem[Gnedin \& Kravtsov(2006)]{2006astro.ph..1401G} Gnedin, N.~Y., \& 
Kravtsov, A.~V.\ 2006, ArXiv Astrophysics e-prints, arXiv:astro-ph/0601401 
 
\bibitem[Grebel \& Gallagher(2004)]{2004ApJ...610L..89G} Grebel, E.~K., \& 
Gallagher, J.~S.\ 2004, ApJL, 610, L89 
 
\bibitem[Heckman et al.(2000)]{2000ApJS..129..493H} Heckman, T.~M., 
Lehnert, M.~D., Strickland, D.~K., \& Armus, L.\ 2000, ApJS, 129, 493 
 
\bibitem[Hensler et al.(2004)]{2004IAUS..217..178H} Hensler, G., 
K{\"o}ppen, J., Pflamm, J., \& Rieschick, A.\ 2004, IAU Symposium, 217, 178 
 
\bibitem[Hernandez et al.(2000)]{2000MNRAS.317..831H} Hernandez, X.,
Gilmore, G., \& Valls-Gabaud, D.\ 2000, MNRAS, 317, 831

\bibitem[Iwamoto et al.(1999)]{1999ApJS..125..439I} Iwamoto, K., Brachwitz, 
F., Nomoto, K., Kishimoto, N., Umeda, H., Hix, W.~R., \& Thielemann, F.-K.\ 
1999, ApJS, 125, 439 
 
\bibitem[Karakas \& Lattanzio(2003)]{2003PASA...20..279K} Karakas, A.~I.~\&
  Lattanzio, J.~C.\ 2003, Publ.~Astron.~Soc.~Aust., 20, 279

\bibitem[Kazantzidis et al.(2004)]{2004ApJ...608..663K} Kazantzidis,
S., Mayer, L., Mastropietro, C., Diemand, J., Stadel, J., \& Moore,
B.\ 2004, ApJ, 608, 663

\bibitem[Kravtsov et al.(2004)]{2004ApJ...609..482K} Kravtsov, A.~V., 
Gnedin, O.~Y., \& Klypin, A.~A.\ 2004, ApJ, 609, 482 
 
\bibitem[Kennicutt(1998)]{1998ApJ...498..541K} Kennicutt, R.~C.\ 1998, 
ApJ, 498, 541 
 
\bibitem[Klypin et al.(1999)]{1999ApJ...522...82K} Klypin, A., Kravtsov, 
A.~V., Valenzuela, O., \& Prada, F.\ 1999, ApJ, 522, 82 
 
\bibitem[Kroupa, Tout, \& Gilmore(1993)]{1993MNRAS.262..545K} Kroupa, P.,
Tout, C.~A., \& Gilmore, G.\ 1993, MNRAS, 262, 545

\bibitem[Lanfranchi \& Matteucci(2003)]{2003MNRAS.345...71L} Lanfranchi, 
G.~A., \& Matteucci, F.\ 2003, MNRAS, 345, 71 
 
\bibitem[Lanfranchi \& Matteucci(2004)]{2004MNRAS.351.1338L} Lanfranchi, 
G.~A., \& Matteucci, F.\ 2004, MNRAS, 351, 1338 
 
\bibitem[Lanfranchi et al.(2006)]{2006MNRAS.365..477L} Lanfranchi, G.~A., 
Matteucci, F., \& Cescutti, G.\ 2006, MNRAS, 365, 477 

\bibitem[Majewski et al.(2000)]{2000LIACo..35..619M} Majewski, S.~R., 
Patterson, R.~J., Dinescu, D.~I., Johnson, W.~Y., Ostheimer, J.~C., Kunkel, 
W.~E., \& Palma, C.\ 2000, Liege International Astrophysical Colloquia, 35, 
619 
 
\bibitem[Marcolini et al.(2003)]{2003MNRAS.345.1329M} Marcolini, A., 
Brighenti, F., \& D'Ercole, A.\ 2003, MNRAS, 345, 1329 
 
\bibitem[Marcolini et al.(2004)]{2004MNRAS.352..363M} Marcolini, A., 
Brighenti, F., \& D'Ercole, A.\ 2004, MNRAS , 352, 363 
 
\bibitem[Marcolini et al.(2006)]{2006astro.ph..2386M} Marcolini, A.,
D'Ercole, A., \& Brighenti, F.\ 2006, ArXiv Astrophysics e-prints,
arXiv:astro-ph/0602386

\bibitem[Martin et al.(2002)]{2002ApJ...574..663M} Martin, C.~L., 
Kobulnicky, H.~A., \& Heckman, T.~M.\ 2002, ApJ, 574, 663 
 
\bibitem[Mashchenko et al.(2005)]{2005ApJ...624..726M} Mashchenko, S., 
Couchman, H.~M.~P., \& Sills, A.\ 2005, ApJ, 624, 726 
 
\bibitem[Matteucci \& Greggio(1986)]{1986A&A...154..279M} Matteucci, F., \& 
Greggio, L.\ 1986, A\&A, 154, 279 
 
\bibitem[Matteucci \& Recchi(2001)]{2001ApJ...558..351M} Matteucci, F., \& 
Recchi, S.\ 2001, ApJ, 558, 351 
 
\bibitem[Mateo(1998)]{1998ARA&A..36..435M} Mateo, M.~L.\ 1998, ARA\&A, 36, 
435 
 
\bibitem[Mayer et al.(2005)]{2005astro.ph..4277M} Mayer, L., Mastropietro, 
C., Wadsley, J., Stadel, J., \& Moore, B.\ 2005, ArXiv Astrophysics 
e-prints, arXiv:astro-ph/0504277 
 
\bibitem[Read et al.(2006)]{2006MNRAS.tmp..153R} Read, J.~I., Wilkinson, 
M.~I., Evans, N.~W., Gilmore, G., \& Kleyna, J.~T.\ 2006, MNRAS, 153 
 
\bibitem[Recchi et al.(2004)]{2004A&A...426...37R} Recchi, S., Matteucci, 
F., D'Ercole, A., \& Tosi, M.\ 2004, A\&A, 426, 37 
 
\bibitem[Recchi et al.(2006)]{2006A&A...445..875R} Recchi, S., Hensler, G., 
Angeretti, L., \& Matteucci, F.\ 2006, A\&A, 445, 875 
 
\bibitem[Ricotti \& Gnedin(2005)]{2005ApJ...629..259R} Ricotti, M., \& 
Gnedin, N.~Y.\ 2005, ApJ, 629, 259 
 
\bibitem[Robertson et al.(2005)]{2005ApJ...632..872R} Robertson, B., 
Bullock, J.~S., Font, A.~S., Johnston, K.~V., \& Hernquist, L.\ 2005, ApJ, 
632, 872 
 
\bibitem[Schaller et al.(1992)]{1992A&AS...96..269S} Schaller, G., 
Schaerer, D., Meynet, G., \& Maeder, A.\ 1992, A\&AS, 96, 269 
 
\bibitem[Shetrone et al.(2003)]{2003AJ....125..684S} Shetrone, M., Venn, 
K.~A., Tolstoy, E., Primas, F., Hill, V., \& Kaufer, A.\ 2003, AJ, 125, 
684 
 
\bibitem[Shu et al.(2005)]{2005ChJAA...5..327S} Shu, C.-G., Mo, H.-J., \& 
Mao, S.-D.\ 2005, Chinese Journal of Astronomy and Astrophysics, 5, 327 
 
\bibitem[Silk(2003)]{2003MNRAS.343..249S} Silk, J.\ 2003, MNRAS, 343, 249 

\bibitem[Smith et al.(2000)]{2000AJ....119.1239S} Smith, V.~V., Suntzeff, 
N.~B., Cunha, K., Gallino, R., Busso, M., Lambert, D.~L., \& Straniero, O.\ 
2000, AJ, 119, 1239 
 
\bibitem[Sumiyoshi et al.(2001)]{2001ApJ...562..880S} Sumiyoshi, K., 
Terasawa, M., Mathews, G.~J., Kajino, T., Yamada, S., \& Suzuki, H.\ 2001, 
ApJ, 562, 880 
 
\bibitem[Thielemann et al.(1986)]{1986A&A...158...17T} Thielemann, F.-K., 
Nomoto, K., \& Yokoi, K.\ 1986, A\&A, 158, 17 
 
\bibitem[Tian et al.(1996)]{1996A&AS..118..503T} Tian, K.~P., van Leeuwen,
F., Zhao, J.~L., \& Su, C.~G.\ 1996, A\&AS, 118, 503

\bibitem[\protect\citeauthoryear{{Timmes}, {Woosley} \& {Weaver}}{{Timmes}
  et~al.}{1995}]{TimmesF_95a}
{Timmes} F.~X.,  {Woosley} S.~E.,    {Weaver} T.~A.,  1995, ApJS, 98, 617

\bibitem[Travaglio et al.(1999)]{1999ApJ...521..691T} Travaglio, C., Galli, 
D., Gallino, R., Busso, M., Ferrini, F., \& Straniero, O.\ 1999, ApJ, 521, 
691 
 
\bibitem[Travaglio et al.(2001)]{2001ApJ...549..346T} Travaglio, C., 
Gallino, R., Busso, M., \& Gratton, R.\ 2001, ApJ, 549, 346 
 
\bibitem[Travaglio et al.(2004)]{2004ApJ...601..864T} Travaglio, C., 
Gallino, R., Arnone, E., Cowan, J., Jordan, F., \& Sneden, C.\ 2004, ApJ, 
601, 864 
 
\bibitem[Tsujimoto et al.(2000)]{2000ApJ...531L..33T} Tsujimoto, T., 
Shigeyama, T., \& Yoshii, Y.\ 2000, ApJL, 531, L33 

\bibitem[Venn et al.(2004)]{2004AJ....128.1177V} Venn, K.~A., Irwin,
M., Shetrone, M.~D., Tout, C.~A., Hill, V., \& Tolstoy, E.\ 2004, AJ,
128, 1177

\bibitem[Wanajo et al.(2003)]{2003ApJ...593..968W} Wanajo, S., Tamamura, 
M., Itoh, N., Nomoto, K., Ishimaru, Y., Beers, T.~C., \& Nozawa, S.\ 2003, 
ApJ, 593, 968 
 
\bibitem[Woosley \& Weaver(1995)]{1995ApJS..101..181W} Woosley, S.~E.~\&
Weaver, T.~A.\ 1995, ApJS, 101, 181

\end{thebibliography}
\end{document}